\documentclass{ws-procs975x65}
\def\sun{{\mbox{$\odot$}}}

\begin{document}

\title{Uncovering galactic and extragalactic planets \\
by gravitational microlensing}

\author{M. Dominik}

\address{University of St Andrews, School of Physics \& Astronomy, \\ North Haugh, St Andrews,
KY16 9SS, United Kingdom
\\E-mail: md35@st-andrews.ac.uk}


\maketitle

\abstracts{
With its planet detection efficiency reaching a maximum for orbital radii
between 1 and 10~AU, microlensing provides a unique sensitivity
to planetary systems similar to our own around 
galactic and even extragalactic stars acting as lenses on observed background stars,
and in particular can detect
terrestrial planets in the habitable zone.
The absence of planetary signals in the 1995--1999 PLANET data 
implies that less than 1/3 of galactic M-dwarfs 
harbour jupiters at orbital radii between
1.5 and 4~AU. If a fraction $f_\mathrm{p}$ of stars is surrounded by a planet,
annual detections of 15--25~$f_\mathrm{p}$~jupiters and 2--3~$f_\mathrm{p}$~earths around galactic stars
would result from PLANET, 
a space-based campaign
would yield 1200~$f_\mathrm{p}$~jupiters and 30~$f_\mathrm{p}$~earths, and
a northern microlensing network 
could detect 15--35~$f_\mathrm{p}$~jupiters and 4--10~$f_\mathrm{p}$~saturns
around M31 stars.}

\section{Planet detection by microlensing}

An {\em unseen planet} of mass $m$ at a projected orbital radius $r_\mathrm{p}$ around
an {\em unseen lens star} of mass $M$ distorts the gravitational field
of its parent star, so that deviations of 1--20$\,$\% (lasting
hours for earths to days for jupiters) to the lightcurve of 
a background star undergoing a microlensing event
(lasting $\sim\,$1~month) can occur.  
The probability for a planet to cause an observable signal ({\em detection efficiency})
is roughly proportional to $\sqrt{q}$
and reaches a maximum near
$r_{\rm p} \sim r_{\rm E}$, where
$r_{\rm E} = \sqrt{2\,R_{\rm S}\,D}$ is the {\em Einstein radius}
of the lens star, with $R_{\rm S} = (2GM)/c^2$ being its
Schwarzschild radius,
$D = D_{\rm L}\,(D_{\rm S}-D_{\rm L})/D_{\rm S}$,
and $D_{\rm S}$ and $D_{\rm L}$ the source or lens distance.

By achieving a galactic length scale
$D \sim 2.5~\mbox{kpc}\;[r_{\rm E}/(2.5~\mbox{AU})]^2$
for typical lens stars, namely M-dwarfs ($M \sim 0.3~M_\odot$), microlensing
becomes sensitive
to planetary systems similar to our own. This condition is
met not only for observations of galactic bulge stars
being lensed by stars in both the galactic disk or the bulge itself
($D_{\rm S} \sim 8.5~\mbox{kpc}$, $D_{\rm L} \sim 6~\mbox{kpc}$),
but also for observations of stars in other galaxies such as M31
lensed by stars
in the same galaxy ($D_{\rm S} \gg D_{\rm S}-D_{\rm L}$, $D \approx D_{\rm L}$).

For a specific mass ratio,
the source star radius sets an upper limit
for the planetary deviation.
With a photometric precision of 1--2\,\% from the ground and
$\sim\,$0.3\,\% from space, microlensing can detect planets with masses
as low as Earth (ground) or even Mars (space).
With individual stars not being resolved in M31 observations,
only events with high peak magnifications on bright source stars (giants)
can be observed at more moderate photometric precision, so that
only planets with masses larger than Saturn can be detected.
 
With its  
signal not depending on the orbital period,
microlensing provides $d = r_\mathrm{p}/r_\mathrm{E}$ as
the only orbital information
and is
blind to inclination and
eccentricity.

\section{Microlensing campaigns on the hunt for planets}

From daily observations of $\sim\,10^{7}$ galactic bulge stars,
OGLE\cite{OGLE} currently announces $\sim\,$500 and
MOA\cite{MOA}
an additional $\sim\,$50 microlensing events 
per year. 
Detection and characterization of planetary deviations
requires dense round-the-clock high-precision
monitoring of events selected in order to maximize
the total planet detection efficiency which can be
achieved by southern telescope networks as 
PLANET\cite{DoEtal},  
MPS\cite{MPS}
or MicroFUN, supplemented by increased sampling
by OGLE and MOA.

A dedicated 1.5m space telescope (GEST)\cite{GEST} would be capable of dense monitoring
of $\sim\,10^{8}$ galactic bulge stars providing $\sim\,$5000 microlensing events per year.

Daily obervations of M31 from at
least four 2m-class northern telescopes\cite{DoRdR} 
including sites used 
by MEGA\cite{MEGA} or AGAPE\cite{SLOTT-AGAPE,AGAPE,POINT-AGAPE} can provide a
sampling of up to 400~events per year every $\sim\,$6~h and more frequently 
on ongoing anomalies.

Table~\ref{camptab} lists and compares the capabilities 
of the different microlensing campaigns (PLANET, GEST,
M31 northern network).

\begin{table}[bt]
\tbl{Capabilities of different microlensing campaigns.
}
{\scriptsize
\begin{tabular}{@{}|l|ccc|@{}}
\hline
\rule{0pt}{2.4ex} & \multicolumn{2}{c}{galactic bulge}  & \\
 & ground & space & \raisebox{1.5ex}[-1.5ex]{M31} \\
\hline
\rule{0pt}{2.7ex}number of source stars & $\sim\,10^{7}$ &
$\sim\,10^{8}$ & $\sim\,10^{10}$\\[0.3ex]
resolution of source stars & resolved/crowded & well-resolved & unresolved \\[0.3ex]
telescope time & dedicated & dedicated & 0.5--2.5~h per night \\[0.3ex]
field of view [sq deg] & 0.004--0.03 & 2 & 0.01--1 \\[0.3ex]
number of fields  &  & & \\[-0.15ex]
monitored during night  &  \raisebox{1.5ex}[-1.5ex]{$\sim 20$} &
\raisebox{1.5ex}[-1.5ex]{$1$} &
\raisebox{1.5ex}[-1.5ex]{1--8} \\[0.3ex]
mean sampling interval & 1.5--2.5~h & 10--15~min& 4--6~h \\[0.3ex]
total event rate $[\mbox{yr}^{-1}]$  & $\sim 300$--600&
$\sim 5000$ & $\sim 150$--400
\\[0.3ex]
useful types & giants & mainly & \\[-0.15ex]
of source stars
& main-sequence stars & main-sequence stars & \raisebox{1.5ex}[-1.5ex]{giants} \\[0.3ex]
useful peak magnifications &
$A_0 \gtrsim 2$ & $A_0 \gtrsim 1.05$ & $A_0 \gtrsim 10$ \\[0.3ex]
rate of useful events $[\mbox{yr}^{-1}]$ & $\sim 75$ &
$\sim 5000$ & $\sim 35$--100 \\[0.3ex]
planet detection efficiency\cite{Covone,GL} & $\sim 20\,$\% (jupiters) & $\sim 25\,$\% (jupiters) &
$\sim 35\,$\% (jupiters)  \\[-0.15ex]
(lensing zone average)
 &
$\sim 1.5\,$\% (earths)  & $\sim 1\,$\% (earths) &
  $\sim 10\,$\% (saturns) \\[0.3ex]
 & 15--25 jupiters &
1200 jupiters & 15--35 jupiters  \\[-0.15ex]
\raisebox{1.5ex}[-1.5ex]{planet probing rate $[\mbox{yr}^{-1}]$} & 2--3 earths &
30 earths & 4--10 saturns  \\[0.3ex]
upper limit on planetary  & 4--7$\,$\% (jupiters) & 0.1$\,$\% (jupiters)
& 3--7$\,$\% (jupiters)\\[-0.15ex]
abundance within 3 years & $\sim\,40\,$\% (earths) & $\sim\,3\,$\% (earths)
& 10--30$\,$\% (saturns)\\[0.3ex]
 & galactic disk  & galactic disk
&
 \\[-0.15ex]
\raisebox{1.5ex}[-1.5ex]{location of parent stars} & galactic bulge & galactic bulge
& \raisebox{1.5ex}[-1.5ex]{M31}
 \\[0.3ex]
extraction of & & & mostly difficult \\[-0.15ex]
planet parameters &\raisebox{1.5ex}[-1.5ex]{fair in many cases}
 & \raisebox{1.5ex}[-1.5ex]{good} & or even impossible \\[0.3ex]
identification of & & for $\sim 33\,$\% &  \\[-0.15ex]
parent stars &\raisebox{1.5ex}[-1.5ex]{no}
 & of the events & \raisebox{1.5ex}[-1.5ex]{no} \\[0.3ex]
isolated and & &  &  \\[-0.15ex]
wide-orbit planets &\raisebox{1.5ex}[-1.5ex]{no}
 & \raisebox{1.5ex}[-1.5ex]{yes} & \raisebox{1.5ex}[-1.5ex]{no} \\
\hline
\end{tabular}\label{camptab} }
\vspace*{13pt}
\end{table}

\section{First results and future prospects}
\begin{figure}[bt]
\centerline{\epsfxsize=2.2in\epsfbox{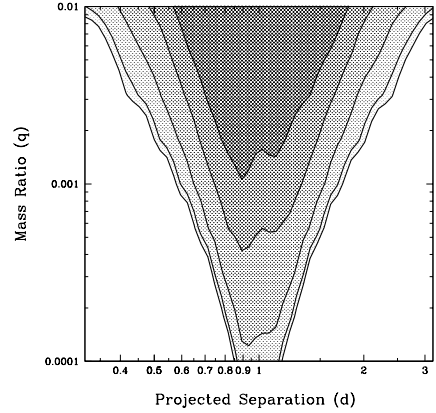}}
\caption{Abundances $f_\mathrm{p}(d,q) =$ ~3/4, 2/3, 1/2, 1/3, and 1/4 (inside to outside)
of planets with $q = m/M$ and $d = r_\mathrm{p}/r_\mathrm{E}$ that are excluded at 95$\,$\% C.L. by 1995--1999 
PLANET data. 
Typically, $M \sim 0.3~M_\sun$ (M-dwarf) and $r_\mathrm{E} \sim 2.5~\mbox{AU}$.}
\label{fig:exclusions}
\end{figure}

The absence of planetary signals in the 1995--1999 PLANET data,
corresponding to probing effectively 9 stars,
yields the upper abundance limits $f_\mathrm{p}(d,q)$ 
shown in Fig.~\ref{fig:exclusions}.\cite{GaudiEtal}
For jupiters ($q \sim 0.003$), $f_\mathrm{p} \lesssim 33\,\%$ for 
$0.6 \leq d \leq 1.6$ ({\em lensing zone}),
corresponding to orbital radii of 1.5--4\,AU,
while a narrower
region around $d \sim 1$
remains sensitive to  
$q \lesssim 10^{-4}$.
Current and future campaigns 
will discover many extra-solar planets within a few years,
drastically reduce their upper abundance limit, or yield a combination of these two scenarios
(see Table~\ref{camptab}).

\begin{thebibliography}{99}
\bibitem{GEST}D.P. Bennett and S.H. Rhie, 
{\it Astrophys.\ J.} {\bf 574}, 985 (2002).
\bibitem{SLOTT-AGAPE}
V. Bozza {\it et al.}, 
{\it Mem. Soc. Astron. Ital.} {\bf  71}, 1113 (2001).
\bibitem{AGAPE}
S. Calchi Novati {\it et al.}, 
{\it Astron. \& Astrophys.} {\bf 381}, 848 (2002).
\bibitem{Covone}G. Covone, R. de Ritis, M. Dominik and A.A. Marino,
{\it Astron. \& Astrophys.} {\bf 357}, 816 (2000).
\bibitem{MEGA}
A.P.S. Crotts, R. Uglesich, G. Gyuk and A.B. Tomaney 
in {\it Gravitational Lensing: Recent Progress and Future Goals},
ed. T.G. Brainerd and C.S. Kochanek (ASP, San Francisco, 2001).
\bibitem{DoRdR} M. Dominik in {\em General Relativity, Cosmology and Gravitational Lensing}, ed. 
G. Marmo, C. Rubano and P. Scudellaro (Bibliopolis, Napoli, 2001).
\bibitem{DoEtal} M. Dominik {\it et al}, {\it Planetary \& Space Science} {\bf 50}, 229 (2002).
\bibitem{GaudiEtal} B.S. Gaudi {\it et al}., {\it Astrophys.\ J.}, {\bf 566}, 463 (2002).
\bibitem{GL}A. Gould and A. Loeb, {\it Astrophys.\ J.} {\bf 396}, 104 (1992).
\bibitem{POINT-AGAPE}
E. Kerins {\it et al.}, 
{\it Mon. Not. Roy. Astron. Soc.} {\bf 323}, 13 (2001).
\bibitem{MPS}S.H. Rhie {\it et al.}, {\it Astrophys.\ J.} {\bf 522}, 1037 (1999).
\bibitem{MOA}T. Sumi {\it et al.}, {\it Astrophys.\ J.} {\bf 591}, 204 (2003).
\bibitem{OGLE}P.R. Wo\'{z}niak {\it et al.}, {\it Acta Astron.} {\bf 51}, 175 (2001).

\end{thebibliography}
\end{document}